\documentclass[aps,preprint]{revtex4}%
\usepackage{amsfonts}
\usepackage{amsmath}
\usepackage{amssymb}
\usepackage{graphicx}%
\providecommand{\U}[1]{\protect\rule{.1in}{.1in}}

\begin{document}
\title{Flow and critical velocity of an imbalanced Fermi gas through an optical potential}
\author{J. Tempere$^{\ast}$}
\affiliation{TFVS, Universiteit Antwerpen, Groenenborgerlaan 171, B-2020 Antwerpen, Belgium}
\begin{abstract}
Optical lattices offer the possibility to investigate the superfluid
properties of both Bose condensates and Fermionic superfluid gases. When a
population imbalance is present in a Fermi mixture, this leads to frustration
of the pairing, and the superfluid properties will be affected. In this
contribution, the influence of imbalance on the flow of a Fermi superfluid
through an optical lattice is investigated. The flow through the lattice is
analysed by taking into account coupling between neighbouring layers of the
optical lattice up to second order in the interlayer tunneling amplitude for
single atoms. The critical velocity of flow through the lattice is shown to
decrease monotonically to zero as the imbalance is increased to 100\%.
Closed-form analytical expressions are given for the tunneling contribution to
the action and for the critical velocity as a function of the binding energy
of pairs in the (quasi) two-dimensional Fermi superfluid and as a function of
the imbalance.

\end{abstract}

\maketitle

\section{Introduction}

In the past several years, ultracold Fermi gases have become a major focus of
research on the interplay between Cooper pairing, strong interactions, and
reduced dimensionality. The appeal of this system stems from the fact that
both the interaction strength and the geometry of the confinement potential
can be controlled very precisely. Feshbach resonances can be applied to tune
the s-wave scattering length over a wide range of positive and negative
values, which allows to investigate the crossover between a regime of
Bose-Einstein condensed (BEC) molecules\cite{FermiBEC} and a
Bardeen-Cooper-Schrieffer (BCS) superfluid\cite{FermiBCS}. Optical lattices
are used to reduce the dimensionality of the system to 1D or 2D or to create a
3D optical lattice\cite{GreinerNAT415,CataliottiSCI293,MIT-optilatt}.
Moreover, since the amount of atoms of each species in a Fermi mixture can
also be controlled accurately, it has become possible to investigate the
effect, on pair formation, of a population imbalance between the pairing
partners\cite{MIT-imbal,RICE-imbal}.

In this contribution, we investigate the case of an imbalanced Fermi
superfluid in a one-dimensional optical potential. In previous work, it was
shown that the gap and number equations can be solved exactly for an
imbalanced two-dimensional Fermi gas\cite{TemperePRB75} -- this corresponds to
a single layer in the optical potential. Here, we will take into account the
coupling between the different layers in the 1D lattice, and derive
expressions for the tunneling amplitude and the critical velocity of the
imbalanced Fermi superfluid through the optical lattice. Tunneling of a
balanced Fermi superfluid through a 1D optical lattice has been studied
experimentally\cite{MIT-optilatt}, as has the imbalanced 3D Fermi
superfluid\cite{MIT-imbal,RICE-imbal}. However, to the best of our knowledge,
tunneling of a imbalanced superfluid has not yet been investigated.

\section{Optical Lattice}

The system that we study in this contribution consists of a Fermi mixture with
unequal amounts of two components, referred to as "spin up" and "spin down".
This is loaded into a one-dimensional optical lattice (in the $z$-direction).
The optical lattice is created by two counterpropagating laser beams with wave
length $\lambda$ and is modelled by a potential energy $V_{0}\sin^{2}\left(
2\pi z/\lambda\right)  $. Under the influence of this lattice, the gas forms a
stack of typically a few hundred quasi-2D layers containing several thousands
of atoms each. The confinement influences the way that interactions are taken
into account. The interaction between atoms within one quasi-2D layer can be
modelled by a 2D-pseudopotential $V=g\delta(\mathbf{r})$ where the strength
$g$ depends on the energy of the scattering atoms through%
\begin{equation}
\frac{1}{g}=\frac{m}{4\hbar^{2}}\left[  i-\frac{\ln\left(  E/E_{b}\right)
}{\pi}\right]  -\int\frac{d^{2}\mathbf{k}}{(2\pi)^{3}}\frac{1}{(\hbar
k)^{2}/m-E+i\varepsilon}.
\end{equation}
Here $m$ is the mass of the atoms, and $E_{b}$ is the energy of the bound
state that always exists in two dimensions\cite{RanderiaPRB41}, given by
$E_{b}=(C\hbar\omega_{L}/\pi) \exp(\sqrt{2\pi} \ell_{L} / a_{s} )$ , with
$a_{s}$ the (3D) s-wave scattering length of the fermionic atoms, $\omega
_{L}=\sqrt{8\pi^{2}V_{0}/\left(  m\lambda^{2}\right)  }$ and $\ell_{L}%
=\sqrt{\hbar/(m\omega_{L})}$ and $C\approx0.915$ (cf. Ref. \cite{PetrovPRA64}).

The goal of this contribution is to take into account the coupling between
adjacent valleys of the optical potential. This coupling is due to tunneling
of individual atoms from layer to layer and is characterised by the tunneling
amplitude\cite{MartikainenPRA68}
\begin{equation}
t=\frac{m\omega_{L}^{2}\lambda^{2}}{8\pi^{2}}\left[  \frac{\pi^{2}}%
{4}-1\right]  e^{-\left(  \lambda/4\ell_{L}\right)  ^{2}}.
\end{equation}

\section{Path-integral approach}

We follow the path-integral approach, as applied by Iskin and Sa de Melo
\cite{IskinPRL97} for an imbalanced gas, and as applied in Ref.
\cite{WoutersPRA70} to the optical lattice. The action functional $S$ for the
fermionic mixture in the optical lattice can be written as a path integral
over the exponential of an action functional, consisting of the contributions
$S_{j}$ of the \emph{individual layers }and the contributions $S_{j,j+1}$ of
\emph{tunneling }between adjacent layers. That is, we write for the action
$S=\sum_{j}\left(  S_{j}+S_{j,j+1}\right)  $, with the single-layer action%
\begin{equation}
S_{j} = {\displaystyle\sum\limits_{k,\sigma}} \bar{\psi}_{k,\sigma}%
^{(j)}\left(  -i\omega_{n}+\mathbf{k}^{2}-\mu_{\sigma}^{(j)}\right)
\psi_{k,\sigma}^{(j)} + g{\displaystyle\sum_{k}} \bar{\psi}_{k,\uparrow}%
^{(j)}\bar{\psi}_{-k,\downarrow}^{(j)}\psi_{-k,\downarrow}^{(j)}%
\psi_{k,\uparrow}^{(j)},
\end{equation}
and the tunneling action%
\begin{equation}
S_{j,j+1}=\sum_{k,\sigma}t\left(  \bar{\psi}_{k,\sigma}^{(j+1)}\psi_{k,\sigma
}^{(j)}+\bar{\psi}_{k,\sigma}^{(j)}\psi_{k,\sigma}^{(j+1)}\right)  .
\end{equation}
Here, we use $k=\{\mathbf{k},\omega_{n}\}$ for the the 2D wave number
\textbf{k} and the Matsubara frequency $\omega_{n},$ and $\sigma$ for the
spin, so that $\bar{\psi}_{k,\sigma}^{(j)}$ and $\psi_{k,\sigma}^{(j)}$
represent the Grassmann variables for the fermionic fields in layer $j$. The
system parameters are the interaction strength $g$, the tunneling amplitude
$t$, and the chemical potentials $\mu_{\sigma}^{(j)}$ that fix the amounts of
spin-up and spin-down particles. We use units $\hbar=2m=1$.

The partition sum is the functional integral of $\exp\{-S\}$ over the
Grassmann variables. The interaction part is decoupled by introducing the
Hubbard-Stratonovic fields $\Delta_{j}$,$\bar{\Delta}_{j}$ in each layer,
after which the functional integral over the Grassmann variables can be taken.
Writing those fields as a function of amplitude and phase, $\Delta
_{j}=\left\vert \Delta_{j}\right\vert e^{i\theta_{j}}$, we find the following
effective action
\begin{equation}
S_{eff}=-\frac{1}{g} \sum_{j} \left\vert \Delta_{j}\right\vert ^{2}%
-\text{Tr}\left\{  \log\left[  -\mathcal{G}_{j,j^{\prime}}^{-1}+\mathcal{T}%
_{j,j^{\prime}}\right]  \right\}
\end{equation}
Here, the trace is taken over $k$, over the spin variables and over the layer
index. The single layer inverse Green's function is%
\begin{equation}
-\mathcal{G}_{j}^{-1}=\delta_{j^{\prime},j}\times\left(
\begin{array}
[c]{cc}%
-i\omega_{n}+\mathbf{k}^{2}-\mu_{\uparrow,j} & \left\vert \Delta
_{j}\right\vert \\
\left\vert \Delta_{j}\right\vert  & -i\omega_{n}-\mathbf{k}^{2}+\mu
_{\downarrow,j}%
\end{array}
\right)
\end{equation}
and the tunneling propagator is%
\begin{equation}
\mathcal{T}_{j,j^{\prime}}=\left[  \delta_{j^{\prime},j+1}+\delta_{j^{\prime
},j-1}\right]  \times\left(
\begin{array}
[c]{cc}%
te^{-i(\theta_{j}-\theta_{j^{\prime}})/2} & 0\\
0 & -te^{i(\theta_{j}-\theta_{j^{\prime}})/2}%
\end{array}
\right)
\end{equation}
Setting $t=0$ we obtain the result for the imbalanced 2D Fermi superfluid,
discussed in Ref. \cite{TemperePRB75}. We write this result as $S_{eff}(t=0)
=\sum_{j}\beta\Omega_{j}$ where $\beta=1/k_{B}T$ is the inverse temperature,
and $\Omega_{j}$ is the BCS free energy for layer $j$. If, on the other hand,
we keep $t>0$ but set $\mu_{\downarrow,j}=\mu_{\uparrow,j}$ we obtain the
results for the balanced Fermi gas in an optical potential, derived in Ref.
\cite{WoutersPRA70}. Here, we keep $t>0$ and investigate $\mu_{\downarrow
,j}\neq\mu_{\uparrow,j}$. We now treat the additional tunneling terms as a
perturbation, and expand up to order $t^{2}$. Diagrammatically, the process we
include in the self-energy of layer $j$ consists of a particle tunneling back
and forth over layer $j+1$ or $j-1$. The result is%
\begin{equation}
S_{eff} = \sum_{j}\left[  \beta\Omega_{j}+\sum_{k} \frac{2t^{2}\cos
(\theta_{j+1}-\theta_{j})\left\vert \Delta_{j}\right\vert \left\vert
\Delta_{j+1}\right\vert } { [(\zeta_{j}-i\omega_{n})^{2}-(E_{\mathbf{k}}%
^{(j)})^{2}] [(\zeta_{j+1}-i\omega_{n})^{2}-(E_{\mathbf{k}}^{(j+1)})^{2}] }
\right]
\end{equation}
where $\sum_{k}$ represents both the sum over Matsubara frequencies and the
integral over $\mathbf{k}.$ Here, $\mu_{j} = (\mu_{\uparrow,j}+\mu
_{\downarrow,j})/2$ is the average chemical potential and $\zeta_{j} =
(\mu_{\uparrow,j}-\mu_{\downarrow,j})/2$ is the difference in chemical
potentials, and
\begin{align}
E_{\mathbf{k}}^{(j)}  &  = \sqrt{(k^{2}-\mu_{j})^{2}+\left\vert \Delta
_{j}\right\vert ^{2}},\\
\xi_{\mathbf{k}}^{(j)}  &  = k^{2}-\mu_{j}.
\end{align}
Note that up to this order of perturbation, the normal particles do not
contribute to tunneling: when we interleave layers of superfluid with layers
of normal gas (for example the excess spin population), the product
$\left\vert \Delta_{j}\right\vert \left\vert \Delta_{j+1}\right\vert $
vanishes. When we assume that no such interleaving takes place and that
moreover $\left\vert \Delta_{j}\right\vert \approx\left\vert \Delta
_{j+1}\right\vert $, the integration over $\mathbf{k}$ and the Matsubara
summation can be performed analytically and results in
\begin{equation}
S_{eff}=\sum_{j}\beta\left[  \Omega_{j}+T_{j+1,j}\cos(\theta_{j+1}-\theta
_{j})\right]  ,
\end{equation}
with%
\begin{equation}
T_{j+1,j}=\frac{t^{2}}{8\pi}\left(  1+\frac{\mu_{j}}{\sqrt{\mu_{j}%
^{2}+\left\vert \Delta_{j}\right\vert ^{2}}}-2\frac{\sqrt{\zeta_{j}%
^{2}-\left\vert \Delta_{j}\right\vert ^{2}}}{\zeta_{j}}\right)  .
\label{tunnelresult}%
\end{equation}
where for $\zeta_{j}<\left\vert \Delta_{j}\right\vert $ the last term is not
present; this case corresponds to the known result for the balanced
gas\cite{WoutersPRA70}.

\section{Josephson regime and critical velocity}

We will investigate the specific case where both the 2D density $n_{\uparrow
}+n_{\downarrow}$ and the imbalance do not change significantly over the
layers. In that case $T_{j+1,j}$, expression (\ref{tunnelresult}), is also not
changing significantly from layer to layer, and we can drop the index $j$ for
$\left\vert \Delta\right\vert ,\mu$ and $\zeta$. The superfluid motion over
the layers is then due to the phase differences over the layers. Using the
results from \cite{TemperePRB75}, we can re-express $T_{j+1,j}$ as a function
of the imbalance $\delta n/n=(n_{\uparrow}-n_{\downarrow})/ (n_{\uparrow
}+n_{\downarrow})$, and of the binding energy per particle $E_{b}/(2E_{F})$,
in units of the Fermi energy $E_{F}=(\hbar k_{F})^{2}/(2m)$ with $k_{F}%
^{2}=2\pi(n_{\uparrow}+n_{\downarrow})$.

Three regimes can be identified. The first regime occurs for large imbalance,
$\delta n/n>E_{b}/(2E_{F})$. Then no superfluidity occurs -- the Fermi mixture
is too imbalanced to support pairing. The second regime is the 'weak pairing'
regime, characterised by $\delta n/n<E_{b}/(2E_{F})$. In the three-dimensional
case, one would refer to this as the BCS regime. Then we have, from Ref.
\cite{TemperePRB75}, the following analytical results for the 2D case:%
\begin{align}
\left\vert \Delta\right\vert ^{2}  &  = 2E_{b} \left[  1-\sqrt{2/E_{b}}
(\delta n / n) \right] \\
\mu &  = 1- (E_{b}/2) \left[  1-\sqrt{2/E_{b}} (\delta n / n) \right] \\
\zeta &  = \sqrt{\left\vert \Delta\right\vert ^{2}+ (\delta n / n)^{2} }%
\end{align}
This results in%
\begin{equation}
T_{j+1,j} = \frac{t^{2}}{8\pi}\left[  \frac{4E_{F}}{2E_{F}+E_{b}-\sqrt
{2E_{F}E_{b}}\frac{\delta n}{n}} -\frac{2\sqrt{E_{F}}\frac{\delta n}{n}}%
{\sqrt{2E_{b}-2\sqrt{2E_{F}E_{b}}\frac{\delta n}{n}+E_{F}\left(  \frac{\delta
n}{n}\right)  ^{2}}}\right]  .
\end{equation}

\begin{figure}[ptb]
\begin{center}
\includegraphics[  width=0.65\linewidth,
keepaspectratio]{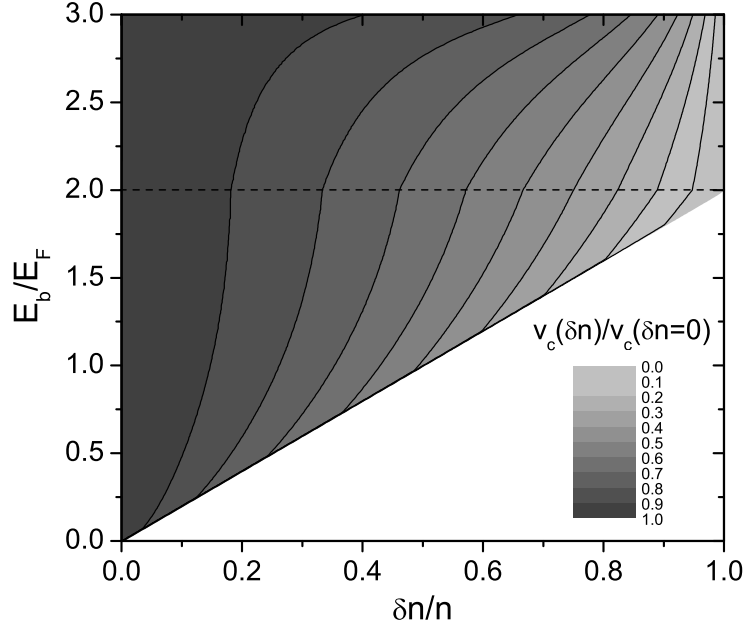}
\end{center}
\caption{ Imbalance between the spin components reduces the critical velocity
of a Fermi superfluid in an optical lattice. Here, the critical velocity
relative to the critical velocity for a balanced gas is shown as a function of
the pair binding energy $E_{b}$ and the imbalance $\delta n$. The dashed line
at $E_{b}=2 E_{F}$ separates the regime of weak pairing from that of strong
pairing. For $E_{b}/E_{F}<2\delta n/n$ (white region) superfluidity is
suppressed.}%
\label{fig1}%
\end{figure}

The third regime is that of strong pairing, characterised by $E_{b}%
/(2E_{F})>1$. In three dimensions this regime might be compared to the BEC
regime. For the strong pairing regime, we find \cite{TemperePRB75}%
\begin{align}
\left\vert \Delta\right\vert ^{2}  &  = 2E_{b}\left(  1-\delta n/n \right) \\
\mu &  = 1-(E_{b}/2) \left(  1- \delta n/n \right) \\
\zeta &  = \sqrt{\left\vert \Delta\right\vert ^{2}+ \left[  2 (\delta n /
n)-\mu\right]  ^{2}}%
\end{align}
which results in
\begin{align}
T_{j+1,j}  &  = \frac{t^{2}}{8\pi}\left[  \frac{4E_{F}}{2E_{F}+E_{b}%
-\sqrt{2E_{F}E_{b}}\frac{\delta n}{n}}\right. \nonumber\\
&  \left.  +2\frac{\left(  2\frac{\delta n}{n}-1\right)  +\frac{E_{b}}%
{2}\left(  1-\frac{\delta n}{n}\right)  }{\sqrt{\left(  2\frac{\delta n}%
{n}-1\right)  ^{2}+\left(  2\frac{\delta n}{n}+1\right)  E_{b}\left(
1-\frac{\delta n}{n}\right)  +\frac{E_{b}^{2}}{4}\left(  1-\frac{\delta n}%
{n}\right)  ^{2}}}\right]
\end{align}
The presence of imbalance reduces $T_{j+1,j}$ from its result without
imbalance. Since the critical velocity can be written as $v_{c}=\lambda
T_{j+1,j}/(\hbar E_{F})$ (see Ref. \cite{TemperePRA72}), we find that that the
imbalance also reduces the critical velocity for flow through the lattice. In
Fig. 1, we show this reduction $v_{c}(\delta n)/v_{c}(\delta n=0),$ as a
function of $E_{b}$ and $\delta n/n$. Note that for $E_{b}/E_{F}<2\delta n/n$
superfluidity is not present. The two regimes of weak pairing and strong
pairing can be distinguished by a kink in the contour lines.

\section{Conclusions}

The theory of flow through the lattice set up Ref. \cite{WoutersPRA70} is
largely unmodified by the presence of imbalance. The main effect is the
reduction of the tunneling coefficient $T_{j,j+1},$ expression
(\ref{tunnelresult}). In this work, we kept $\delta n$ constant and assume
that $\left\vert \Delta\right\vert ,\mu,\zeta$ vary slowly from one lattice
site to another. Under these assumptions, we have derived closed analytical
formulae for the tunneling contributions (and thus the critical velocity) for
flow of an imbalanced superfluid through an optical lattice.

The assumption that $\left\vert \Delta\right\vert ,\mu,\zeta$ vary slowly is
good unless we have interleaving of normal gas and superfluid gas layers. If
this is the case tunneling is suppressed because we take the optical potential
to be deep enough such that the normal gas is pinned and flow is only due to
phase coherence of the pair condensate. The assumption of constant $\delta n$
is related; it also relies on the fact that there is no phase separation. This
will be much more difficult to satisfy in practice, as coexistence may only be
achievable at finite temperatures. In practice, at low temperature, there will
be phase separation in the layer and the tunneling will only take place in the
region of superfluid phase, and will be suppressed in the normal layer around
it. This dynamical interplay between pinned normal state, and a phase
separated superfluid are outside the scope of this paper, which relies on the
possibility to create (albeit dynamically) an imbalanced superfluid.

\begin{acknowledgments}
Discussions with J. T. Devreese, M. Wouters and D. Lemmens are gratefully
acknowledged. This research has been supported financially by the FWO-V
projects Nos. G.0356.06, G.0115.06, G.0435.03, and the GOA BOF UA 2000 UA.
J.T. gratefully acknowledges support of the Special Research Fund of the
University of Antwerp, BOF NOI UA 2004.
\end{acknowledgments}

\end{document}